\documentclass[12pt]{article}
\usepackage[letterpaper,margin=1in]{geometry}
\usepackage{newtxtext,newtxmath}
\usepackage{microtype}
\usepackage[normalem]{ulem}
\usepackage{graphicx}
\usepackage{xcolor}
\usepackage{soul}
\usepackage[hidelinks]{hyperref}
\usepackage{url}
\usepackage[numbers,sort&compress]{natbib}
\usepackage[left]{lineno}
\usepackage{setspace}
\usepackage[font=small,labelfont=bf]{caption}

\newcommand{\scititle}{Sign-switchable and tunable macroscopic curvature of liquid jets by microscopic laser-induced cavitation}

\title{\Large\bfseries\scititle}

\author{
Sourabh Singh$^{1}$,
S.~Sree~Harsha$^{1}$,
Prashant Kumar Singh$^{1,*}$\\[1em]
$^{1}$Tata Institute of Fundamental Research Hyderabad,\\
36/P Gopanpally, Hyderabad 500046, Telangana, India\\[0.5em]
$^{*}$Correspondence: \href{mailto:pks@tifrh.res.in}{pks@tifrh.res.in}
}

\date{}

\begin{document}


\maketitle

\begin{abstract}
Localized forcing of liquid jets can trigger Rayleigh--Plateau breakup at low perturbation amplitudes or explosive fragmentation at high amplitudes, while controlled bending in the intermediate regime has remained unexplored. We show that femtosecond-laser-induced cavitation, seeded at the interface of a $100~\mu$m-diameter water jet, produces a selectable curvature in $0.4$--$1$~mm long segments. Cavitation rupturing at the liquid--air interface drives fast outward jetting and induces a slower recoil on the opposite boundary, which evolves on the capillary--inertial timescale and results in curving of the liquid segment. The curvature is governed by three laser parameters. Off-axis laser illumination on the jet shifts the refractive caustics and redistributes energy deposition between the front and rear interfaces. The resulting recoil imbalance determines the curvature sign, with a straight configuration in the balanced case. Laser energy offers additional control over the recoil impulse via plasma defocusing and nonlinear absorption in the liquid jet, where the response is bounded by axisymmetric deformation at low energy and explosive vaporization at high energy. Repetition rate sets the spacing between successive cavitation events and hence the length of the bent segment. Curvature extends across the segment for lengths of 6--10 jet diameters; shorter segments undergo Rayleigh--Plateau breakup, whereas longer ones bend mainly near their retracting ends. Our results offer a platform for optofluidic control of switchable and tunable macroscopic curvature in flowing liquid jets.
\end{abstract}


\section*{Introduction}

Curved shapes in flowing liquid columns can be observed when viscous ropes coil under gravity~\cite{Coiling_PRL_2004, Coiling_PRE_2008, Coiling_NatCom_2026}, charged jets whip in electric fields~\cite{Whipping_PNAS_2014, Whipping_PRL_2014}, viscous columns buckle under axial compression~\cite{Buckling_PRL_2010, Buckling_PRL_2012}, and acoustic radiation pressure deforms them~\cite{Lonzaga2007, Fang2024, Marr2001}. Optofluidics offers a different means of actuation~\cite{Laser_actuation_fluid}, with the possibility of non-contact access, micrometre-scale spatial localization, and ultrafast energy deposition. However, established approaches to optical manipulation rely either on radiation pressure, which acts directly on free interfaces but generally produces only nanometric displacements because of the small momentum of photons~\cite{Ashkin1973, Casner_Delville2001}, or on absorption-mediated thermocapillary flows~\cite{Garnier2003, Schmidt2025}, which evolve over thermal-diffusion timescales. Femtosecond laser pulses provide an alternative by inducing nonlinear optical breakdown in liquids~\cite{optical_breakdown_Chang1987, optical_breakdown_Mazur1997, optical_breakdown_Mazur2002, optical_breakdown_PRL_2002}. The resulting plasma launches shock waves~\cite{Shock_Vogel1996, PhysRevFluids, LasoGarcia2024}, drives material ablation~\cite{ablation_droplet_Chang1987, ablation_droplet_Lindinger2004, ablation_droplet_Klein2015}, and initiates cavitation~\cite{Energy_cavitation, Cavitation_Vassholz2021, Cheng2025, Patrascioiu2014, Feng2026, Blake1987}. Of these processes, cavitation offers a particularly effective route to mechanical actuation: the rapid expansion and subsequent collapse of a cavitation bubble~\cite{Bubble_osciilation_2010}, together with the associated liquid jetting, impart a localized mechanical impulse~\cite{Wang2025, Shangguan1997, Poulain2015, Ren2022} whose recoil drives liquid motion on the capillary timescale characteristic of thin jets~\cite{Singh_Rayleigh_2026}.

These processes are widely used to manipulate liquids and solids, but their effects remain largely confined to the irradiated object or focal volume. In droplets, ablation and cavitation primarily drive the deformation, propulsion, or fragmentation of the drop itself~\cite{optical_breakdown_Chang1987, ablation_droplet_Chang1987, ablation_droplet_Lindinger2004, droplet_cavitation_PRL2006, ablation_droplet_Klein2015, Xu2025, Jagadale2024, Meijer2022, Liu2021, Krivokorytov2017, Kurilovich2018, Cheng2024, ablation_droplet_Singh2026}. In bulk transparent solids, femtosecond pulses induce permanent modifications localized to the focal volume, a principle exploited in laser micromachining~\cite{Laser_writing_PRL_2003, Laser_writing_natphot_2008, Laser_writing_review_2021, Laser_writing_nature2026}. A liquid jet offers a distinct possibility: it is deformable, unlike a solid, yet axially extended, unlike a droplet. Its hydrodynamic continuity allows a localized perturbation to influence length scales far exceeding both the focal spot and the jet diameter~\cite{Eggers_RPP2008}. Yet laser-driven liquid-jet studies have so far focused on two distinct outcomes: high-fluence excitation, leading to explosive vaporization, ablation and jet fragmentation~\cite{Gpa_Shock_Nat2016, jet_shock_Nano2020, jet_shock_2022}, and lower-fluence excitation, producing predominantly axisymmetric perturbations that seed Rayleigh--Plateau breakup~\cite{Rayleigh_CWlaser_2021, Singh_Rayleigh_2026}. Neither regime addresses the possibility of bending a flowing jet while preserving its continuity. Accessing this intermediate regime requires a delicate balance between the deposited energy, its location, and the length of the perturbed segment, against capillary instability, end retraction and pinch-off~\cite{Eggers_RPP2008}.

Here, we show that multi-kilohertz femtosecond laser pulses impart switchable, tunable macroscopic curvature to a water jet. Although each cavitation event is confined to a region comparable to the jet diameter ($D \approx 100\,\mu$m), it bends a segment several times longer ($0.4$--$1$\,mm). Time-resolved imaging shows that cavitation-driven jetting is too fast for the column to follow, and that the column instead responds to the slow net recoil on its capillary--inertial timescale, bending away from the more strongly driven interface. Three laser parameters set the outcome. First, the pulse energy must lie within an intermediate range: weak cavitation produces predominantly axisymmetric perturbations~\cite{Singh_Rayleigh_2026}, excessive energy vaporizes and fragments the jet, and only in between does the recoil bend a segment while it remains intact. Second, the repetition rate sets the aspect ratio $L/D$ of the segment between successive pulses, with $L = U/f$, and curvature spans the whole segment only for $L/D \approx 6$--10, between capillary-stable perturbations and Rayleigh--Plateau breakup at short lengths and end retraction and pinch-off at long lengths. Third, because the focal spot is much smaller than the jet, the cylindrical lensing of the jet~\cite{Caustic_Book, Cylinder_focusOL, Singh2026Caustic} allows the transverse impact parameter to move the dominant interaction from the front to the rear interface, switching the curvature sign. After excitation, the bend persists and evolves through retraction of the segment ends until the segment collapses into a droplet. This control over both the sign and magnitude of curvature offers a way to shape flowing liquids dynamically, with potential use as adaptive optical elements~\cite{adaptive_lens_APL2003, adaptive_lens_APL2004, adaptive_lens_Nature2006, Lopez2008, Nguyen2010} and curved plasma-mirror targets~\cite{Vincenti2019, Vincenti2023, Jeong2025}.


\section*{Results}

\subsection*{Bounded window for laser-induced jet curvature}

\begin{figure}[h!]
	\centering
	\includegraphics[width=1\textwidth]{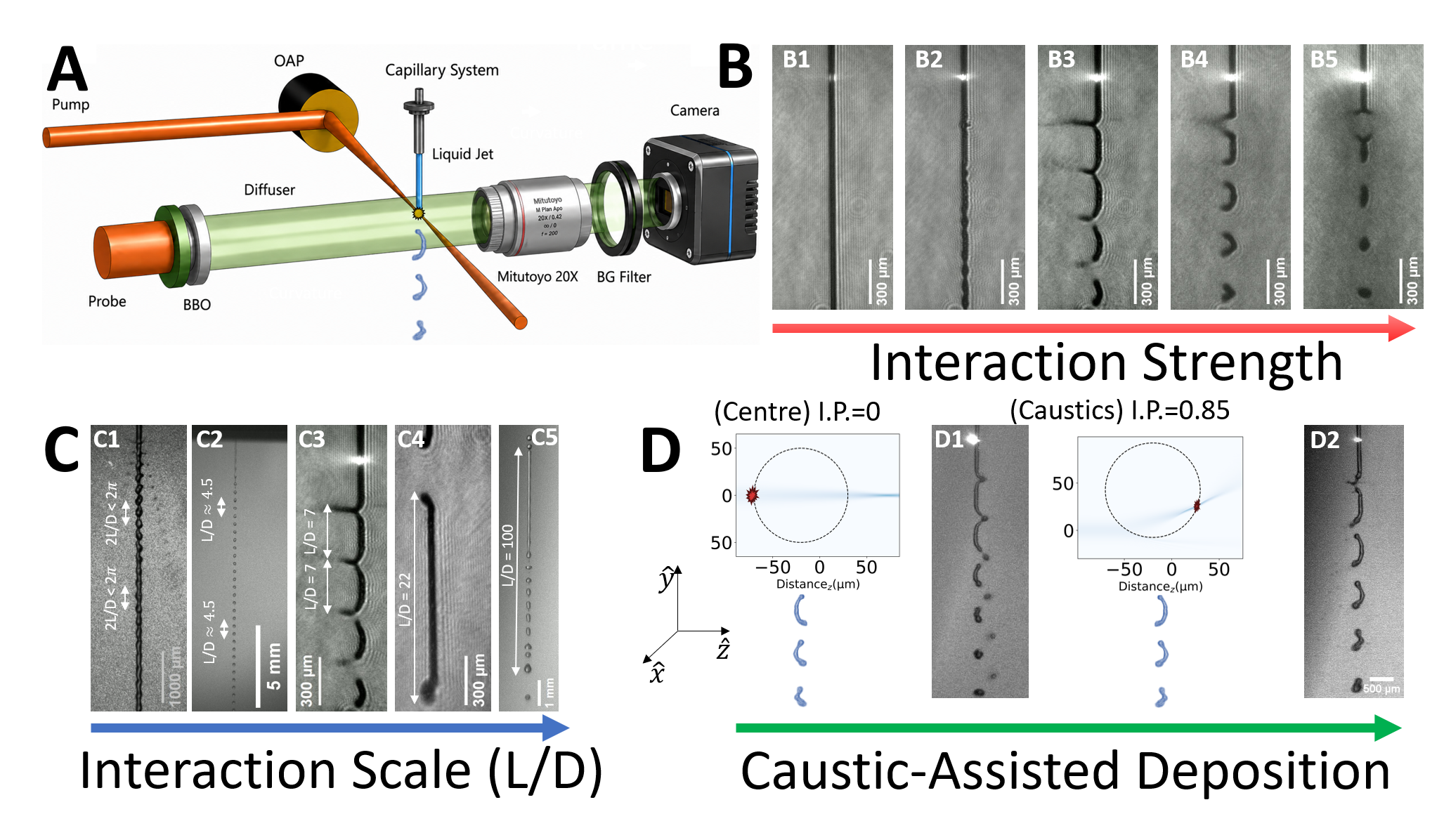}
	\caption{\textbf{Parameter space for laser-driven curvature of liquid jet.}
	\textbf{(A)} Experimental geometry: multi-kilohertz femtosecond pulses ($1/e^2$ radius $\approx 12\,\mu\mathrm{m}$) are focused perpendicular to the flow onto a water jet ($D \approx 100\,\mu\mathrm{m}$, $U = 10\text{--}15\,\mathrm{m\,s^{-1}}$). \textbf{(B)} Dependence on laser pulse energy: (\textbf{B1}) below optical breakdown threshold; (\textbf{B2}) axisymmetric bulging; (\textbf{B3}--\textbf{B4}) asymmetric cavitation recoil and nonlocal bending, with increasing curvature; (\textbf{B5}) excessive ablation and loss of liquid continuity. \textbf{(C)} Dependence on segment aspect ratio $L/D$, controlled by laser repetition rate: (\textbf{C1}) $L/D \lesssim \pi$: below the Rayleigh--Plateau cutoff, no growth; (\textbf{C2}) $L/D \sim 4.5$, Rayleigh--Plateau breakup; (\textbf{C3}) $L/D \approx 6\text{--}10$, extended bending; (\textbf{C4}) $L/D \gtrsim 20$, endpoint-localized bending; (\textbf{C5}) $L/D \gtrsim 50$, fragmentation by endpoint pinching. \textbf{(D)} Control of curvature sign through transverse impact position: on-axis (\textbf{D1}) and off-axis (\textbf{D2}) irradiation shift the breakdown location between opposite interfaces, reversing the cavitation recoil and the resulting curvature.}
	\label{experimentalsetup}
\end{figure}

Imposing curvature on a macroscopic segment of a liquid jet through localized laser-induced cavitation depends on three control parameters: the pulse energy, which sets the strength of the cavitation recoil; the repetition rate, which sets the segment aspect ratio $L/D$; and the transverse impact position, which sets the internal caustic geometry and hence the breakdown location (Fig.~\ref{experimentalsetup}). We first identify the energy window that breaks the axisymmetry of the jet without fragmenting it (Fig.~\ref{experimentalsetup}B). The interaction is mediated by nonlinear optical breakdown, requiring intensities exceeding $\sim10^{13}$~W~cm$^{-2}$ to seed plasma formation via multiphoton, tunnelling, and cascade ionization~\cite{Optical_breakdown_Vogel, Optical_breakdown_Water}. Below this threshold ($E_{\text{laser}} < 2.0~\mu$J), the jet remains unperturbed (Fig.~\ref{experimentalsetup}B1). Just above the breakdown threshold, the response remains nearly axisymmetric: the jet develops a local bulge that seeds Rayleigh--Plateau growth downstream~\cite{Rayleigh_CWlaser_2021, Rayleigh_nslaser_2021, Singh_Rayleigh_2026} (Fig.~\ref{experimentalsetup}B2). As the pulse energy increases to ~40 $\mu$J, the recoil becomes strong enough to displace liquid transversely. Because the segment remains intact, this local transverse displacement is accompanied by deformation of the adjoining liquid ~\cite{deGennes2004,Lin2003}, producing a curved segment that extends well beyond the cavitation region (Fig.~\ref{experimentalsetup}B3,B4). At excessive energies ($\gtrsim 100~\mu$J), explosive vaporization and ablation dominate, fragmenting the segment and destroying the liquid connection required for bending~\cite{Gpa_Shock_Nat2016, jet_shock_Nano2020, jet_shock_2022} (Fig.~\ref{experimentalsetup}B5).

With the energy window established, we next examine the hydrodynamic length-scale constraints. The laser repetition rate $f$ sets the segment length $L = U/f$, and hence its aspect ratio $L/D$ (Fig.~\ref{experimentalsetup}C). Unlike whipping~\cite{Whipping_PNAS_2014, Whipping_PRL_2014}, coiling~\cite{Coiling_PRL_2004, Coiling_PRE_2008, Coiling_NatCom_2026}, or buckling~\cite{Buckling_PRL_2010, Buckling_PRL_2012}, where the forcing (electric stresses, gravity, or axial compression) is distributed along the filament, here the segment is bounded by two consecutive cavitation sites, and forcing is applied only at its ends. This restricts the range of segment lengths over which the curvature can extend across the entire segment. For the shortest spacings ($L/D \lesssim \pi$), the forcing wavelength lies below the Rayleigh--Plateau cutoff ($kR > 1$, with $k = 2\pi/L$), so the imposed perturbations are capillary-stable and do not grow; the jet remains continuous, with neither droplet formation nor global bending (Fig.~\ref{experimentalsetup}C1). At $L/D \approx 4.5$, the forcing length coincides with the fastest-growing Rayleigh--Plateau wavelength ($\lambda_{\mathrm{opt}} \approx 4.5D$)~\cite{Plateau1857, Rayleigh1878, Eggers_RPP2008}, and capillary breakup into droplets dominates before a sustained curvature can develop (Fig.~\ref{experimentalsetup}C2). For long segments ($L/D \gtrsim 20$), the recoil still bends the segment locally near its ends, but this bending does not propagate along the segment. The free ends retract under surface tension at approximately the Taylor--Culick speed, $\sqrt{\sigma/\rho R} \approx 1$~m~s$^{-1}$, collecting liquid into end bulbs, rounded swellings thicker than the jet, that recede along an essentially straight midsection (Fig.~\ref{experimentalsetup}C4). For $L/D \gtrsim 50$, the segment fragments into droplets by capillary end pinching (Fig.~\ref{experimentalsetup}C5), as expected for our water jet, whose Ohnesorge number $Oh = \mu/\sqrt{\rho\sigma D} \approx 0.012$ is too low for viscous damping to suppress end pinching~\cite{jet_breakup_Schulkes_1996, jet_breakup_BASARAN_2004, jet_breakup_PRL2012}. Only within an intermediate window ($L/D \approx 6$--$10$) does the curvature span the entire segment (Fig.~\ref{experimentalsetup}C3). Beyond this window, the deformation remains confined to the segment ends, which retract into end bulbs and eventually pinch off, leaving the midsection straight (Fig.~\ref{experimentalsetup}C4,C5).

Beyond controlling the amplitude of the deformation, we exploit the jet's intrinsic optical geometry to deterministically control the sign of the curvature (Fig.~\ref{experimentalsetup}D). Because the jet diameter ($\sim100~\mu$m) greatly exceeds the laser focal spot ($\sim20~\mu$m), the cylindrical liquid--air interface acts as a dielectric lens. We quantify the irradiation geometry via the impact parameter, $\mathrm{I.P.}=b/R$, where $b$ is the lateral offset between the laser axis and the jet center, and $R$ is the jet radius. Tuning $\mathrm{I.P.}$ reconfigures the internal caustic intensity distribution. Under asymmetric irradiation ($\mathrm{I.P.}=0.85 \pm0.1$), caustic-assisted focusing produces a tight line focus at the distal interior boundary (the rear surface). This localizes the optical breakdown and cavitation recoil at the back of the jet, driving a forward recoil and producing a negative curvature sign (e.g., concave, Fig.~\ref{experimentalsetup}D2). Conversely, shifting to symmetric irradiation ($\mathrm{I.P.}=0$) translates the breakdown to the front boundary, reversing the direction of the recoil and the resulting bending moment to produce the positive curvature sign (e.g., convex, Fig.~\ref{experimentalsetup}D1). 

\subsection*{Recoil-driven origin of curvature}

\begin{figure}[h!]
	\centering
	\includegraphics[width=\linewidth]{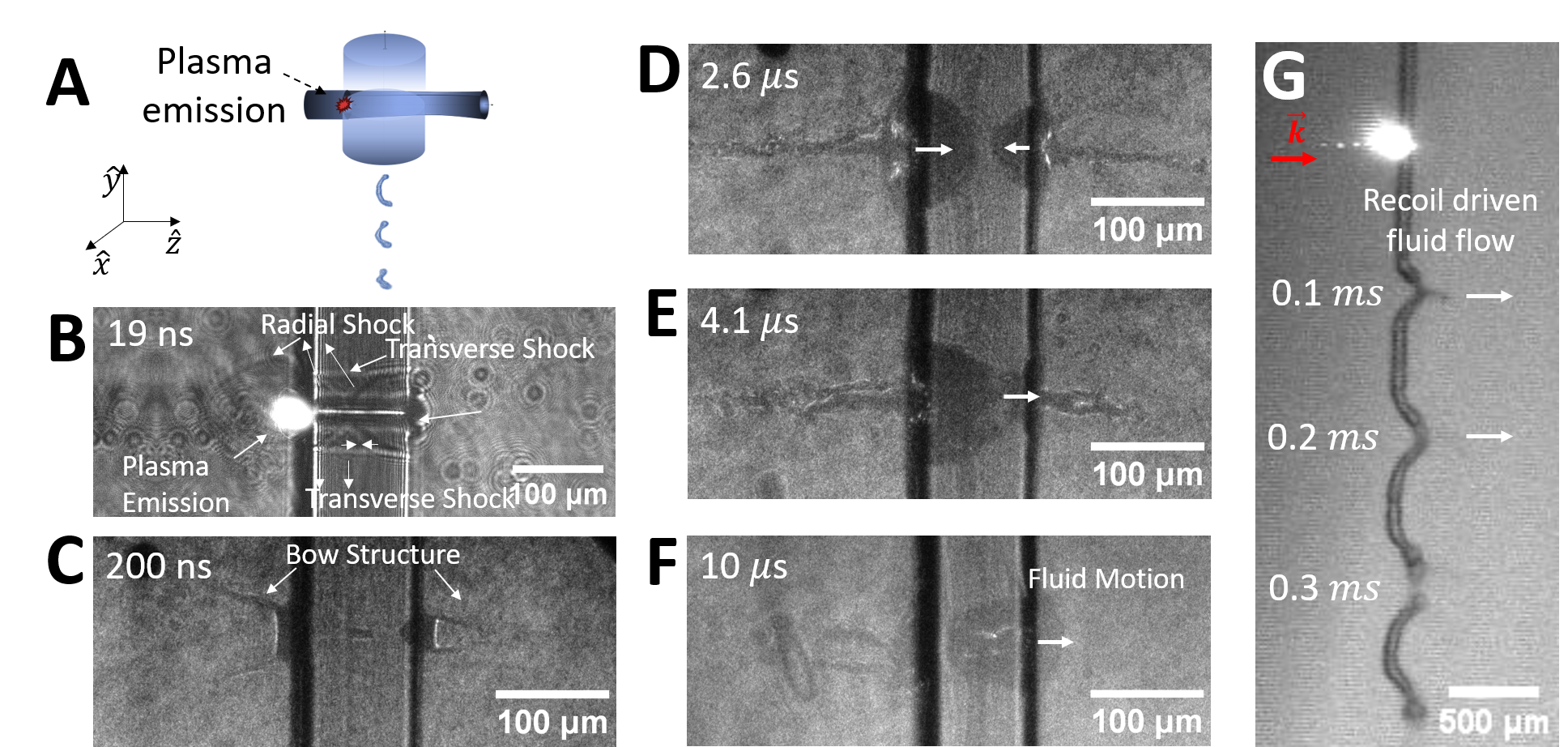}
	\caption{\textbf{Time-resolved dynamics of curvature formation under centred irradiation ($\mathrm{I.P.} = 0 \pm 0.1$).}
\textbf{(A)} Schematic of centred irradiation: breakdown occurs at the front interface, setting the sign of the resulting curvature.
\textbf{(B)} At 19\,ns, a plasma channel spans the jet along the beam path, brightest at the front interface, and a shock propagates outward perpendicular to the beam.
\textbf{(C)} At $\sim$200\,ns, liquid is ablated outward from both interfaces, much more strongly from the front.
\textbf{(D)} At 2.6\,$\mu$s, cavitation ruptures both interfaces, producing high-speed jetting from front and rear, with the front jetting much stronger.
\textbf{(E)} At 4.1\,$\mu$s, the front-dominated jetting continues on microsecond timescales, far shorter than the capillary time of the jet ($\tau_c \approx 42\,\mu$s), so the column cannot follow it.
\textbf{(F)} At $\sim$10\,$\mu$s, as jetting subsides, the net recoil from the imbalance between front and rear jetting drives a slow bulk motion of the column toward the rear, on the capillary--inertial timescale of the jet.
\textbf{(G)} At 0.1\,ms ($E = 70\,\mu$J), this recoil-driven motion bends the segment away from the incident beam, opposite to the side of strongest interaction.}
	\label{centerdynamics}
\end{figure}

To identify how a localized optical impulse produces curvature, we follow the interaction across its characteristic timescales using time-resolved pump--probe imaging. A femtosecond pulse deposits its energy within the pulse duration, yet the liquid column can respond only on its capillary--inertial timescale, $\tau_c = \sqrt{\rho R^3/\sigma} \approx 42~\mu$s, some eight orders of magnitude later. Cavitation bridges this gap by converting the deposited energy into a net transverse momentum. Figure~\ref{centerdynamics} resolves this conversion for centred irradiation ($\mathrm{I.P.} = 0 \pm 0.1$).

For pulse energies above $11~\mu$J, the energy deposition is strongly asymmetric. At 19~ns, the brightest plasma emission occurs at the front interface (Fig.~\ref{centerdynamics}B), indicating that most of the pulse energy is absorbed there, where the energy density is therefore highest. Part of the pulse is transmitted into the jet, forming a plasma channel along the beam path from which a shock propagates outward, perpendicular to the beam. By $\sim$200~ns, the ablated liquid forms bow-shaped structures at both interfaces~\cite{Vogel2003}, with the front structure substantially larger, reflecting the higher energy density at the front (Fig.~\ref{centerdynamics}C).

Although breakdown occurs along the entire plasma channel (Fig.~\ref{centerdynamics}B), cavitation requires a higher local energy density, which is reached at the two interfaces. At the front, the pulse arrives before any absorption within the jet, giving the highest irradiance and the strongest coupling, as the bright emission in Fig.~\ref{centerdynamics}B shows. At the rear, the transmitted light is concentrated by the cylindrical lensing of the jet (Fig.~\ref{fig:frontbackintensity}). Cavitation therefore nucleates at both interfaces on microsecond timescales~\cite{Optical_breakdown_Vogel, Robert2007}, more strongly at the front. Bubbles nucleated at the interfaces of a $100~\mu$m jet cannot expand freely as in an unbounded liquid. They instead rupture the interfaces and drive high-speed jetting from both sides, with the front jetting dominant (Fig.~\ref{centerdynamics}D)~\cite{Blake1987, Robert2007, Patrascioiu2014}. The jetting lasts several microseconds (Fig.~\ref{centerdynamics}E), shorter than $\tau_c$, so the column cannot follow it. What survives the jetting is its momentum. The imbalance between front and rear jetting leaves the column with a net recoil directed toward the rear, which, as the jetting subsides, drives a slow bulk motion at the irradiation site over tens of microseconds, on the capillary--inertial timescale (Fig.~\ref{centerdynamics}F). It is this slow motion, not the jetting itself, that displaces the column transversely.

The dynamics in Fig.~\ref{centerdynamics}B--F describe a single pulse acting at a single site, which displaces the column only locally. Curving a segment requires the same transverse recoil at both of its ends, and this is provided by a multi-kilohertz pulse train acting on the flowing jet. Successive pulses, separated by $1/f$, irradiate fresh sites a distance $L = U/f$ apart, so each segment is bounded by two irradiation sites and receives the same rearward recoil at both ends. The subsequent evolution of this segment produces its curvature, which points away from the side of strongest interaction. By 0.1~ms, the segment is markedly bent while remaining intact (Fig.~\ref{centerdynamics}G, $E = 70~\mu$J), and the deformation persists as it is carried downstream up to $\sim$0.3~ms.

\subsection*{Caustic control of the curvature sign}

The recoil mechanism implies that the curvature can be reversed, making the bending bidirectional. Since the column follows the slow recoil rather than the jetting (Fig.~\ref{centerdynamics}), the segment bends away from the interface where cavitation nucleates, which is the interface receiving the higher energy density. Shifting cavitation from one interface to the other should therefore reverse the curvature. We achieve this by translating the laser focus across the jet: offsetting the beam from the jet axis lowers the irradiance at the front interface and, through the cylindrical lensing of the jet, concentrates the transmitted light at the rear, shifting cavitation from front to rear~\cite{Singh2026Caustic, Singh_Rayleigh_2026}. We take the curvature as positive when the segment bends along the beam and negative when it bends toward the laser. 

\begin{figure}[!htbp]
	\centering
	\includegraphics[width=1\linewidth]{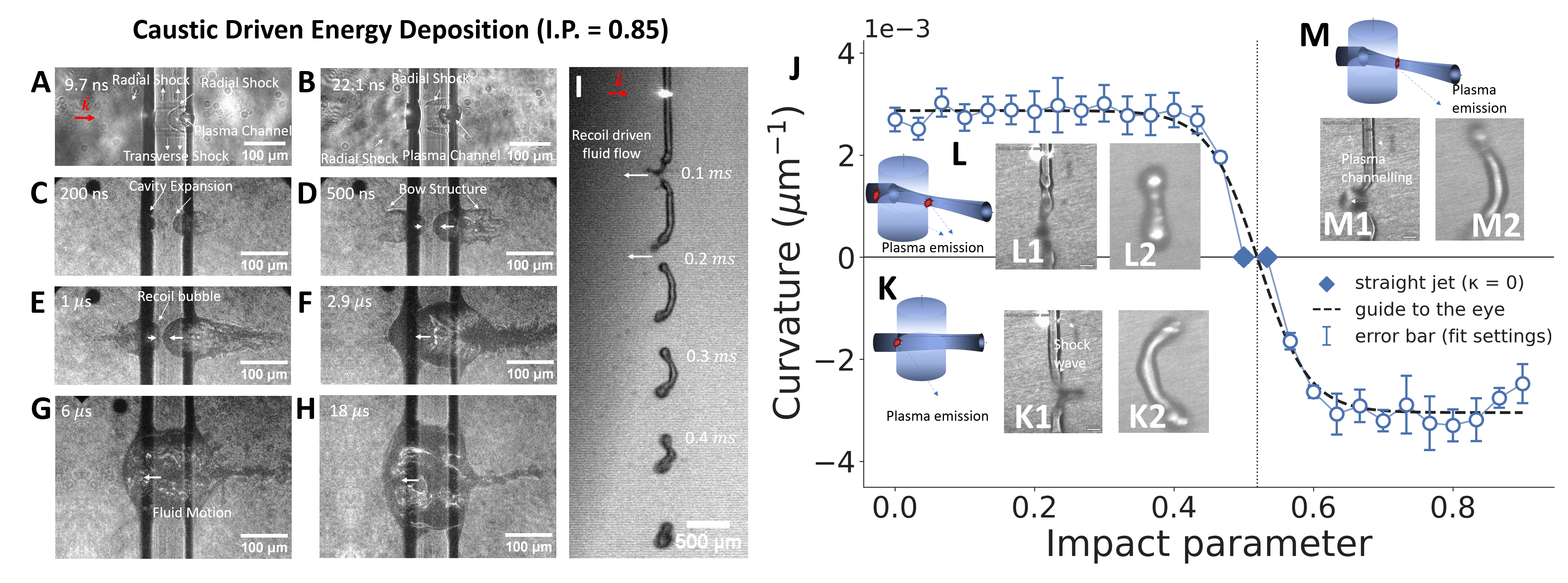}
    \caption{\textbf{Caustic control of the curvature sign.}
    \textbf{(A)--(I)} Dynamics under off-centre irradiation ($\mathrm{I.P.} = 0.85 \pm 0.1$).
    \textbf{(A)} 9.7\,ns: plasma forms at both interfaces and along a channel across the jet, launching shocks from both sides, with the rear plasma region and shock much stronger.
    \textbf{(B)} 22.1\,ns: the plasma region at the rear interface dominates.
    \textbf{(C)} $\sim$200\,ns: bow-shaped ablation structures form at both interfaces, larger at the rear.
    \textbf{(D)} $\sim$500\,ns: the bow-shaped structures continue to develop, remaining larger at the rear.
    \textbf{(E)} 1\,$\mu$s: cavitation ruptures the interfaces, with the dominant jetting from the rear.
    \textbf{(F)} 2.9\,$\mu$s: the cavitation region is displaced toward the front.
    \textbf{(G, H)} 6--18\,$\mu$s: liquid is transported toward the front.
    \textbf{(I)} 0.1\,ms: the segment bends toward the laser, opposite to the centred case.
    \textbf{(J)} Curvature versus impact parameter, showing two plateaus of $\kappa \approx \pm 3 \times 10^{-3}~\mu$m$^{-1}$ separated by a sharp sign reversal near $\mathrm{I.P.} \approx 0.52$, where the jet remains straight (diamonds). The dashed line is a guide to the eye, and error bars reflect variation with fit settings.
    \textbf{(K)--(M)} Schematics of plasma formation at $\mathrm{I.P.} \approx 0$ (K), $0.5$ (L), and $0.85$ (M), with shadowgraphs during the interaction (1) and of the resulting segment (2).
    \textbf{(K1)} Front-dominated plasma formation, with the shock wave propagating through the jet. \textbf{(K2)} The resulting segment bends along the beam ($\kappa > 0$).
    \textbf{(L1)} Plasma forms at both interfaces with comparable strength. \textbf{(L2)} The segment remains straight ($\kappa \approx 0$).
    \textbf{(M1)} Rear-dominated plasma formation, with plasma channelling through the jet. \textbf{(M2)} The resulting segment bends toward the laser ($\kappa < 0$).}
    \label{offcenterdynamics}
\end{figure} 

Under centred irradiation ($\mathrm{I.P.} = 0$), the beam meets the front interface at normal incidence, where its irradiance is highest, and the front dominates the deposition (Fig.~\ref{centerdynamics}). As the impact parameter increases, the beam meets the front interface at increasingly oblique incidence. The geometric projection spreads the focal spot over a larger area of the curved surface and lowers the front irradiance, while refraction and the cylindrical caustic concentrate the transmitted light near the rear interface (Figs.~\ref{fig:frontbackintensity} and \ref{ip_focal})~\cite{Singh2026Caustic}. At $\mathrm{I.P.} = 0.85 \pm 0.1$ and pulse energies of about 3--12~$\mu$J, the deposition is therefore dominated by the rear interface.

The sequence of plasma formation, shock emission, cavitation, jetting, and recoil is the same as under centred irradiation, but with the roles of the front and rear interfaces reversed (Fig.~\ref{offcenterdynamics}A--I). At 9.7~ns, plasma forms at both interfaces and along a channel across the jet, launching shocks from both sides, but the rear plasma region is much stronger and its shock markedly more pronounced (Fig.~\ref{offcenterdynamics}A). By 22.1~ns, the plasma region at the rear interface clearly dominates (Fig.~\ref{offcenterdynamics}B). The bow-shaped ablation structures at $\sim$200 and $\sim$500~ns are larger at the rear (Fig.~\ref{offcenterdynamics}C,D), and cavitation, strongest at the rear, ruptures the interfaces and drives the dominant jetting from the rear by 1~$\mu$s (Fig.~\ref{offcenterdynamics}E). The cavitation region is displaced toward the front by 2.9~$\mu$s (Fig.~\ref{offcenterdynamics}F), and once the jetting subsides, the net recoil drives a slow bulk motion of the column toward the front over 6--18~$\mu$s (Fig.~\ref{offcenterdynamics}G,H). Under the pulse train, each segment receives this recoil at both ends and bends toward the laser by 0.1~ms (Fig.~\ref{offcenterdynamics}I), opposite to the centred case. This rear-dominated response persists across the pulse-energy range under off-centre irradiation, where caustic focusing increases energy transmission towards the rear interface and consequently enhances energy deposition there (Supplementary Fig.~\ref{energy_curvature_ablation}).

Varying the impact parameter continuously reveals a sharp transition between the two regimes (Fig.~\ref{offcenterdynamics}J and Supplementary Videos 1, 2, 3 and 4). For $\mathrm{I.P.} \lesssim 0.45$, plasma forms predominantly at the front interface and the curvature is positive and nearly constant, $\kappa \approx +3 \times 10^{-3}~\mu$m$^{-1}$ (Fig.~\ref{offcenterdynamics}K). Near $\mathrm{I.P.} \approx 0.5$, plasma forms at both interfaces with comparable strength (Fig.~\ref{offcenterdynamics}L1), the opposing recoils nearly cancel, and the segment remains straight (Fig.~\ref{offcenterdynamics}L2; see also Fig.~\ref{shock_bubble}). For $\mathrm{I.P.} \gtrsim 0.6$, the rear interface dominates and the curvature reverses to a nearly constant $\kappa \approx -3 \times 10^{-3}~\mu$m$^{-1}$ (Fig.~\ref{offcenterdynamics}M). The curvature thus switches between two plateaus of equal magnitude and opposite sign within $\Delta\mathrm{I.P.} \approx 0.1$. By selecting the interface at which cavitation nucleates, the impact parameter sets the sign of the curvature with little effect on its magnitude.


\subsection*{Laser pulse-energy-driven front--rear intensity crossover}

\begin{figure}[h!]
	\centering
	\includegraphics[width=\linewidth]{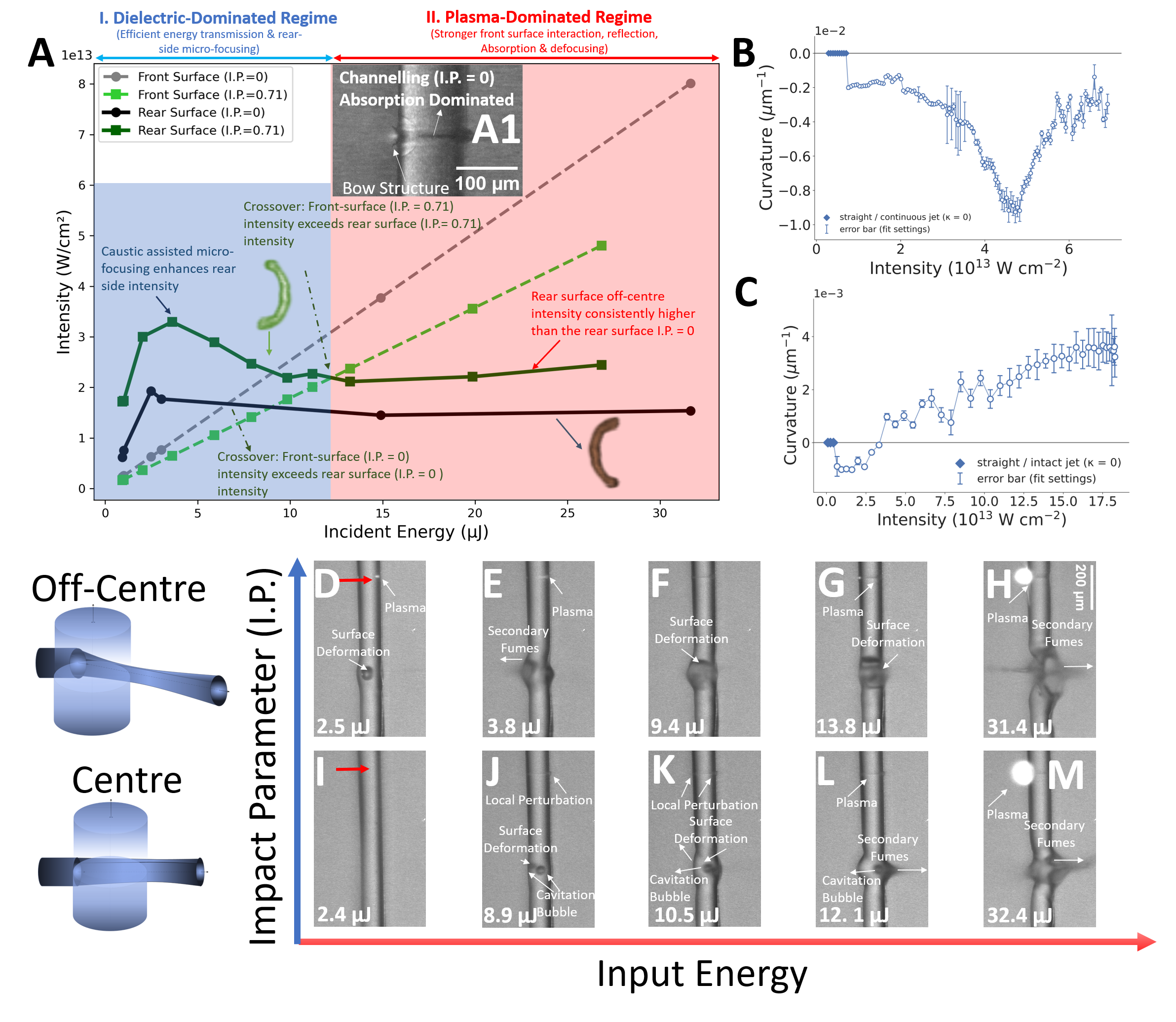}
	\caption{\textbf{Intensity crossover and curvature reversal with pulse energy.} 
    \textbf{(A)} Effective front-surface (dashed) and rear-surface (solid) intensities versus incident pulse energy for $\mathrm{I.P.} = 0$ and $\mathrm{I.P.} = 0.71$, with the dielectric-dominated (blue) and plasma-dominated (red) regimes indicated. \textbf{(A1)} Plasma channel inside the jet at $\mathrm{I.P.} = 0$.
    \textbf{(B)} Curvature versus effective incident intensity $I$ (defined in Methods) for a $50~\mu$m jet, showing the complete window: onset near $0.8 \times 10^{13}$~W~cm$^{-2}$, a maximum near $4.7 \times 10^{13}$~W~cm$^{-2}$, and a decrease as explosive vaporization sets in.
    \textbf{(C)} Curvature versus effective incident intensity $I$ for the $100~\mu$m jet at $\mathrm{I.P.} = 0.8$, showing a small negative excursion, a sign change near $3 \times 10^{13}$~W~cm$^{-2}$, and positive curvature approaching saturation.
    \textbf{(D--H)} Shadowgraphs $\sim$40\,$\mu$s after irradiation at $\mathrm{I.P.} = 0.71\pm 0.1$ for pulse energies of 2.5--31.4\,$\mu$J. Full series in Supplementary Fig.~\ref{energy_ip_curvature} 
    \textbf{(I--M)} Shadowgraphs $\sim$40\,$\mu$s after irradiation at $\mathrm{I.P.} = 0$ for pulse energies of 2.4--32.4\,$\mu$J.}
	\label{fig:frontbackintensity}
\end{figure}

Since the segment bends away from the more strongly driven interface, we compare the effective front- and rear-surface intensities to identify how the interaction asymmetry changes with pulse energy. Figure~\ref{fig:frontbackintensity}A compares the effective intensities at the front and rear interfaces. The front-surface intensity follows from the incident energy and the projected focal spot, and therefore increases linearly with energy, with a smaller slope at $\mathrm{I.P.} = 0.71$ because oblique incidence enlarges the projected spot. The rear-surface intensity is obtained from the measured transmitted energy and the measured energy-dependent rear focal spot~\cite{Singh2026Caustic}.

At low pulse energies, the jet behaves as a dielectric cylindrical lens (Fig.~\ref{ip_focal}). Apart from Fresnel losses of about 4\%, the light reaches the rear interface, where cylindrical focusing concentrates it into a smaller spot. The rear intensity therefore exceeds the front intensity, and more so at $\mathrm{I.P.} = 0.71$, where caustic focusing is strongest~\cite{Singh2026Caustic}. The rear intensity peaks near 2~$\mu$J at $\mathrm{I.P.} = 0$ and near 4~$\mu$J at $\mathrm{I.P.} = 0.71$. At higher energies, optical breakdown forms a dense plasma channel inside the jet (Fig.~\ref{fig:frontbackintensity}A1), and the rear intensity falls for two reasons. Absorption along the channel reduces the energy reaching the rear interface, while the plasma reflectivity remains modest at $\approx$10--15\%~\cite{Singh2026Caustic}. In addition, plasma defocusing and related nonlinear propagation effects broaden the rear focal spot~\cite{Feit1974, Couairon2007, Singh2026Caustic}. The front intensity continues to rise linearly, so the two cross at $\approx$6~$\mu$J for $\mathrm{I.P.} = 0$ and at $\approx$12--13~$\mu$J for $\mathrm{I.P.} = 0.71$, beyond which the effective front-surface intensity is higher. Throughout the measured range, the off-centre rear intensity remains above the head-on rear intensity, so caustic focusing retains an advantage even in the plasma-dominated regime.

The shadowgraphs, recorded $\sim$40~$\mu$s after irradiation, reveal at which interface cavitation and jetting dominate across this energy range, and hence the direction of the recoil (full series in Supplementary Fig.~\ref{energy_ip_curvature}). Under off-centre irradiation, a weak surface deformation appears at 2.5~$\mu$J (Fig.~\ref{fig:frontbackintensity}D). Between 3.8 and 9.4~$\mu$J, within the rear-dominated regime, the perturbations originate at the rear interface (Fig.~\ref{fig:frontbackintensity}E,F). Near the crossover, at 13.8~$\mu$J, perturbations of comparable strength appear at both interfaces (Fig.~\ref{fig:frontbackintensity}G), and at 31.4~$\mu$J the front interface dominates (Fig.~\ref{fig:frontbackintensity}H). Under centred irradiation, no perturbation is observed at 2.4~$\mu$J (Fig.~\ref{fig:frontbackintensity}I). The rear-dominated regime at $\mathrm{I.P.} = 0$ is thus confined to energies below the cavitation threshold, so rear-side jetting does not occur. Cavitation and surface deformation appear between 8.9 and 10.5~$\mu$J (Fig.~\ref{fig:frontbackintensity}J,K), and from 12.1~$\mu$J the front interface dominates (Fig.~\ref{fig:frontbackintensity}L,M). The change in the dominant interfacial response is consistent with a change in recoil direction, whereas appreciable segment curvature appears only at higher energies. At $\mathrm{I.P.} = 0.8$, the curvature remains zero below $\approx 0.5 \times 10^{13}$~W~cm$^{-2}$, becomes slightly negative ($\kappa \approx -1 \times 10^{-3}~\mu$m$^{-1}$) up to $\approx 2.5 \times 10^{13}$~W~cm$^{-2}$, and changes sign near $3 \times 10^{13}$~W~cm$^{-2}$ (Fig.~\ref{fig:frontbackintensity}C and Supplementary Videos 6), close to the intensity at which the front and rear intensities cross in Fig.~\ref{fig:frontbackintensity}A. Above this intensity, the curvature is positive and grows to $\approx 3.5 \times 10^{-3}~\mu$m$^{-1}$ at $\approx 1.5 \times 10^{14}$~W~cm$^{-2}$, approaching saturation without decreasing up to $1.8 \times 10^{14}$~W~cm$^{-2}$.

A thinner, $50~\mu$m diameter jet captures the complete window (Fig.~\ref{fig:frontbackintensity}B and Supplementary Videos 5). Curvature appears abruptly above $\approx 0.8 \times 10^{13}$~W~cm$^{-2}$, its magnitude grows to a maximum of $|\kappa| \approx 9 \times 10^{-3}~\mu$m$^{-1}$ near $4.7 \times 10^{13}$~W~cm$^{-2}$, and then falls to $\approx 3 \times 10^{-3}~\mu$m$^{-1}$ at higher intensities, as explosive vaporization of the irradiated region destroys the continuity needed for the column to follow the recoil. The thinner jet thus bends more strongly but over a much narrower range, with its curvature peaking at an intensity nearly four times lower than that at which the $100~\mu$m jet is still bending more strongly. These data quantify the sequence of regimes shown in Fig.~\ref{experimentalsetup}B. Intensity therefore sets the magnitude of the curvature within a window whose upper bound decreases for thinner jets, and, through the front--rear crossover, its sign at a fixed impact parameter.

\subsection*{Curvature evolution during capillary retraction}

\begin{figure}[h!]
	\centering
	\includegraphics[width=\linewidth]{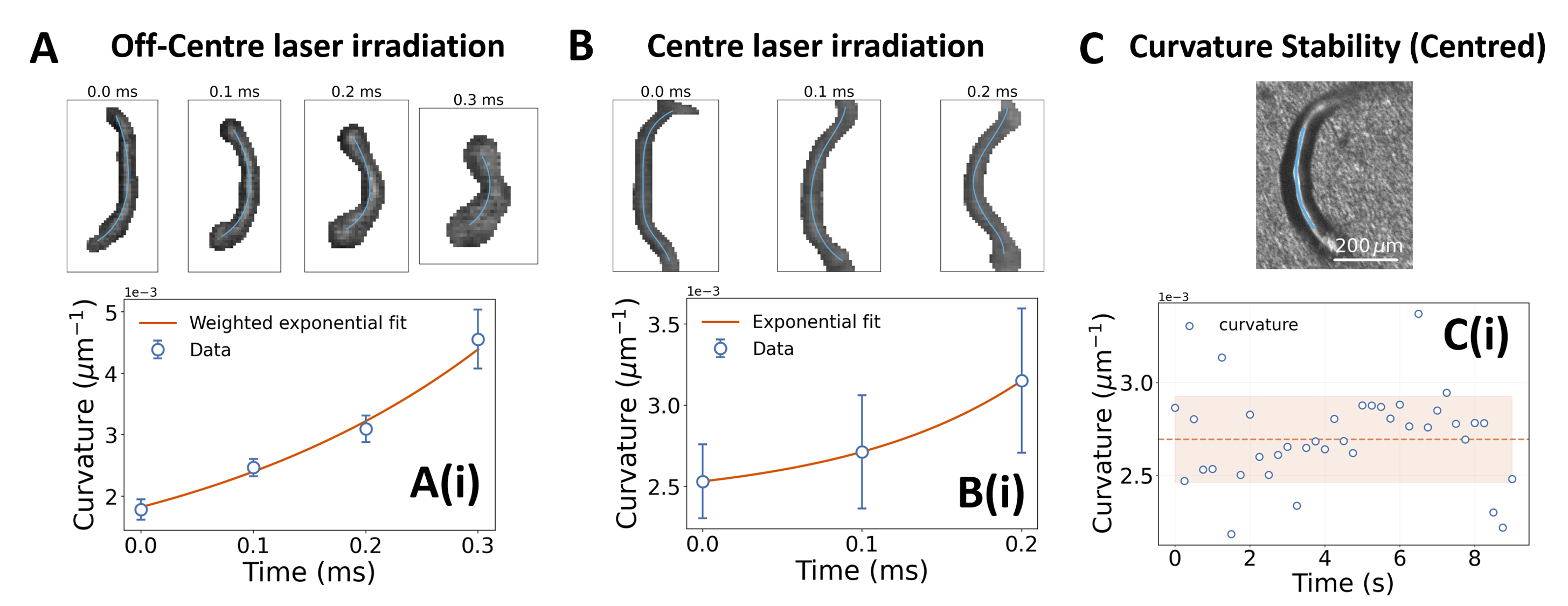}
    \caption{\textbf{Evolution and stability of the laser-imprinted curvature.}
    \textbf{(A)} Off-centre irradiation ($\mathrm{I.P.} = 0.71$): shadowgraphs of a bent segment from 0 to 0.3~ms, with fitted centrelines (blue).
    \textbf{(A(i))} Measured curvature versus time, with a weighted exponential fit.
    \textbf{(B)} Centred irradiation ($\mathrm{I.P.} = 0$): shadowgraphs from 0 to 0.2~ms of a segment bent with the opposite curvature sign, with fitted centrelines.
    \textbf{(B(i))} Measured curvature versus time, with an exponential fit.
    \textbf{(C)} Shadowgraph of a bent segment under centred irradiation, with its fitted centreline.
    \textbf{(C(i))} Curvature from repeated measurements over $\approx 9$~s. The dashed line marks the mean, and the shaded band spans one standard deviation.
    Time zero corresponds to the first analysed frame.}
    \label{fig:jetcurvaturea}
\end{figure}

Once imposed, the curvature evolves without further optical forcing. We track isolated bent segments ($L/D \approx 10$) as they flow downstream, fit their centrelines, and extract the mean curvature $\kappa$ at successive times (Fig.~\ref{fig:jetcurvaturea}A,B and Supplementary Section~\ref{sec:curvature_method}). Under off-centre irradiation ($\mathrm{I.P.} = 0.71$), the curvature increases steadily from $\kappa \approx 1.8 \times 10^{-3}$ to $4.6 \times 10^{-3}~\mu$m$^{-1}$ within 0.3~ms (Fig.~\ref{fig:jetcurvaturea}A(i)), while the segment shortens, its ends round into bulbs, and it contracts toward a compact drop (Fig.~\ref{fig:jetcurvaturea}A). Under centred irradiation, the segment bends with the opposite sign, and its curvature increases more modestly, from $\approx 2.5 \times 10^{-3}$ to $3.2 \times 10^{-3}~\mu$m$^{-1}$ over 0.2~ms, with larger uncertainty (Fig.~\ref{fig:jetcurvaturea}B,B(i)).

The imposed bend persists because bending a liquid column at fixed length costs no surface energy. By Pappus's theorem, a bent cylindrical segment of radius $R$ has the same lateral area, $2\pi RL$, and volume, $\pi R^2 L$, as a straight one (Supplementary Section~\ref{sec:bent_segment}), so surface tension exerts no restoring force toward a straight shape. The evolution is instead dominated by retraction of the free ends, which shortens the segment and gathers its liquid into end bulbs that eventually merge into a single drop, as also observed for straight segments (Supplementary Fig.~\ref{shock_bubble}F). For a straight segment, we measure a retraction speed of $\approx 0.4$~m~s$^{-1}$ per end, about one third of the capillary speed $\sqrt{\sigma/\rho R} \approx 1.2$~m~s$^{-1}$~\cite{Taylor1959, Culick1960, Keller1983}. Bending also makes the Laplace pressure slightly higher on the outer side of the bend than on the inner side, which could contribute to the tightening. Because the stronger irradiation required for bending removes more liquid, however, segment length and volume vary from case to case, and our data cannot separate this contribution from end retraction (Supplementary Section~\ref{sec:bent_segment}). For $L/D \approx 10$, the segment contracts into one drop rather than fragmenting by end pinching (Fig.~\ref{experimentalsetup}C4,C5).

In a paraxial geometric picture, the measured curvature corresponds to an equivalent focal length $|f_{\mathrm{eff}}| = 1/(2|\kappa|)$ for reflection in the plane of the bend, a quantity derived from $\kappa$ rather than measured optically. It decreases from $\approx 280$ to $\approx 110~\mu$m under off-centre irradiation and from $\approx 200$ to $\approx 160~\mu$m under centred irradiation, with opposite signs for the two geometries. The circular cross-section of the jet adds a much stronger curvature in the orthogonal plane, so such a reflector would be astigmatic. The laser thus selects the sign and initial value of the curvature, which then evolves until the segment collapses. Repeated measurements under centred irradiation show that the imposed curvature is reproducible over $\approx 9$~s, with mean $\bar{\kappa} = 2.7 \times 10^{-3}~\mu$m$^{-1}$ and standard deviation $\delta\kappa = 2.3 \times 10^{-4}~\mu$m$^{-1}$, a relative fluctuation of $\approx 9\%$ (Fig.~\ref{fig:jetcurvaturea}C,C(i)).


\section*{Discussion}
Our results show that a localized femtosecond-laser impulse can impose a selectable macroscopic curvature on a flowing liquid jet. The mechanism relies on the separation between cavitation-driven jetting, which begins on sub-microsecond timescales, and the much slower capillary--inertial response of the liquid column, $\tau_c \approx 42~\mu$s. The jet bends in response to the net recoil, away from the more strongly driven interface. Thus, the laser sets the initial impulse, while capillary dynamics govern the subsequent deformation. The curvature is controlled by three parameters. The impact parameter selects its sign through cylindrical lensing, producing two plateaus of opposite curvature separated by a straight configuration where the opposing recoils balance. Pulse energy controls the recoil strength and shifts this balance through the breakdown-mediated redistribution of energy between the front and rear interfaces. The repetition rate sets the segment length, with distributed bending confined to \(L/D\approx6\)--10. Shorter segments undergo capillary breakup, while longer ones deform predominantly near their ends. The imposed bend persists until capillary retraction collapses the segment into a droplet. The mechanism may also operate in other transparent liquids, although the accessible response will depend on optical breakdown, energy coupling, viscosity, and surface tension. Contactless curvature control offers a route to adaptive liquid optics~\cite{adaptive_lens_APL2003, adaptive_lens_APL2004, adaptive_lens_Nature2006, Lopez2008, Nguyen2010} and curved liquid targets for laser--plasma interactions~\cite{Vincenti2019, Vincenti2023, Jeong2025, Mondal2024}.

\section*{Methods}

\subsection*{Multi-kHz Femtosecond Laser system}
Experiments were performed at TIFR Hyderabad using a commercial Yb:KGW laser system (Pharos, Light Conversion Ltd.) delivering 175-fs pulses at 1034~nm, with pulse energies up to 2~mJ and repetition rates up to 200~kHz. The beam, 7~mm in diameter, was focused by a gold-coated off-axis parabolic mirror with focal length 100~mm. The focal spot, characterized with a 20$\times$ Mitutoyo NIR objective (NA 0.42), had $1/e^{2}$ radii $w_x = 12.9~\mu$m and $w_y = 11.4~\mu$m. Throughout this work, we characterize the incident irradiation by the effective intensity $I = E/(\pi w_x w_y \tau)$, where $E$ is the pulse energy and $\tau$ the pulse duration, which equals $1.27 \times 10^{12}$~W~cm$^{-2}$ for $E = 1~\mu$J. For Gaussian spatial and temporal profiles, with $\tau$ the full width at half maximum, the corresponding peak intensity is $I_{\mathrm{peak}} = (4\sqrt{\ln 2}/\pi^{3/2})\,E/(w_x w_y \tau) \approx 1.88\,I$, or $2.4 \times 10^{12}$~W~cm$^{-2}$ at $E = 1~\mu$J. These values refer to the incident beam in the absence of the jet. The intensities at the front and rear interfaces, which include the projected spot and the cylindrical lensing of the jet, are given in Fig.~\ref{fig:frontbackintensity}A.

\subsection*{Liquid-jet generation}
All measurements were carried out under ambient conditions using deionized water. A pressurized reservoir operating at 1--4~bar supplied a vertically propagating laminar jet with velocities of 10--20~m~s$^{-1}$, generated through a laser-micromachined metallic nozzle. The interaction region was placed 2--3~mm downstream of the nozzle, where the jet remained continuous and hydrodynamically stable, well upstream of Rayleigh–Plateau breakup. Alignment between the jet and laser focus was adjusted using manual three-axis translation stages.

\subsection*{Time-resolved imaging}
The laser--liquid interaction was imaged by pump--probe shadowgraphy (8-bit CMOS camera) with a 517~nm probe, frequency-doubled from the 1034~nm fundamental in a BBO crystal. BG39 filters blocked plasma self-emission, and a diffuser homogenized the probe illumination. A delay line in the probe arm resolved the nanosecond plasma and shock dynamics, and the shot-to-shot separation of the high-repetition-rate laser on the continuously flowing jet captured the evolution from hundreds of microseconds to milliseconds. Cavitation dynamics were imaged separately with an ICCD camera (20$\times$ Mitutoyo NIR objective, 4--50~ns gate) under continuous-wave red diode-laser illumination ($\lambda \approx 630$--680~nm), with a bandpass filter blocking plasma self-emission and residual pump light. A photodiode trigger and a digital delay generator synchronized the camera to the pump pulse. Pulse energies were kept below 30~$\mu$J to avoid ICCD saturation by intense plasma emission.




\section*{Acknowledgments}
\paragraph*{Funding:}
The authors acknowledge support of the Department of Atomic Energy, Government of India, under Project Identification No. RTI4007.
\paragraph*{Author contributions:} P.K.S. and S.S. proposed the concept and initiated the project. P.K.S., S.S., and S.S.H. contributed to the experimental setup. With input from P.K.S. and S.S.H., S.S. led the data analysis. P.K.S. and S.S. wrote the manuscript. P.K.S. provided overall project guidance. 
\paragraph*{Competing interests:}
The authors declare no competing interests.
\paragraph*{Data and materials availability:}
All data needed for evaluations of conclusions in the paper are present in the main and/or in the Supplementary sections. Any additional data related to this paper can be requested from the authors.

\newpage

\renewcommand{\thefigure}{S\arabic{figure}}
\renewcommand{\thetable}{S\arabic{table}}
\renewcommand{\theequation}{S\arabic{equation}}
\renewcommand{\thepage}{S\arabic{page}}
\setcounter{figure}{0}
\setcounter{table}{0}
\setcounter{equation}{0}
\setcounter{page}{1} 


\clearpage
\begin{center}
{\LARGE\bfseries Supplementary Information}

\vspace{0.8cm}

{\Large\bfseries \scititle}

\vspace{0.8cm}

Sourabh Singh$^{1}$,
S.~Sree~Harsha$^{1}$,
Prashant Kumar Singh$^{1,*}$

\vspace{0.4cm}

{\small
$^{1}$Tata Institute of Fundamental Research Hyderabad\\
36/P Gopanpally, Hyderabad 500046, Telangana, India
}

\vspace{0.3cm}

{\small
*Correspondence: \texttt{pks@tifrh.res.in}
}

\end{center}
\vspace{1cm}
\setcounter{section}{0}
\section*{Supplementary Text}
\renewcommand{\thesubsection}{S\arabic{subsection}}


\subsection{Optical characterization: focal-spot calibration and energy dependence}
\label{focal_absorption}
\label{sec:caustic_profiles}

\begin{figure}[h!]
	\centering
	\includegraphics[width=0.8\textwidth]{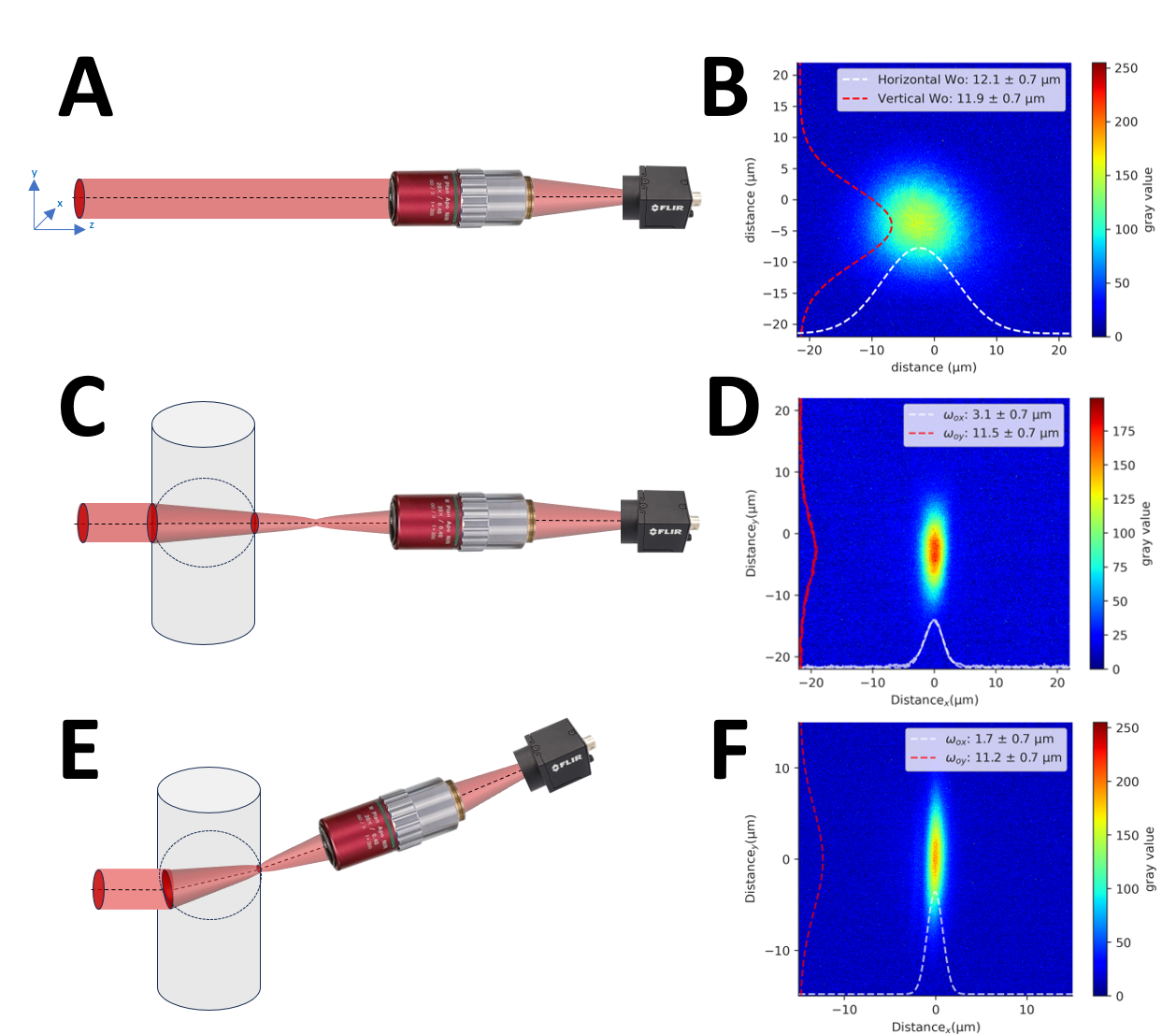}
    \caption{\textbf{Experimental configurations and corresponding measured focal intensity profiles.} 
    \textbf{(A)} Schematic of the setup used to measure the laser focal spot without the liquid jet. 
    \textbf{(B)} Measured focal intensity distribution without cylindrical lensing by the jet. 
    \textbf{(C)} Configuration at an impact parameter (I.P.) of 0, with the laser focus positioned outside the liquid jet. 
    \textbf{(D)} Measured focal intensity distribution at the rear surface of the jet for I.P. = 0. 
    \textbf{(E)} Configuration at I.P. = 0.71, illustrating caustic focusing toward the rear surface. 
    \textbf{(F)} Measured focal intensity distribution at the rear surface for I.P. = 0.71, showing the stronger concentration of the transmitted light.}
	\label{ip_focal}
\end{figure}

To quantify the irradiation geometry, we define the impact parameter $\mathrm{I.P.} = b/R$, where $b$ is the lateral offset between the laser axis and the jet centre and $R$ is the jet radius. Most reproducibility measurements were performed at $\mathrm{I.P.} = 0.71$, where the reflected beam propagates nearly perpendicular to the incident beam, which simplifies alignment of the jet relative to the reflected beam and enables reliable characterization of the impact-parameter-dependent response. The measured focal intensity profiles for the different irradiation geometries are shown in Fig.~\ref{ip_focal}.

The reference focal spot, measured without the liquid jet (Fig.~\ref{ip_focal}A), has a nearly circular Gaussian profile with $1/e^2$ waist radii $\omega_{0x} = 12.1~\mu$m and $\omega_{0y} = 11.9~\mu$m (Fig.~\ref{ip_focal}B). Under centred irradiation ($\mathrm{I.P.} = 0$, Fig.~\ref{ip_focal}C), cylindrical lensing by the jet compresses the beam predominantly along $x$, producing an elliptical focal spot with $\omega_{0x} = 3.1~\mu$m and $\omega_{0y} = 11.5~\mu$m (Fig.~\ref{ip_focal}D). Under off-centre irradiation ($\mathrm{I.P.} = 0.71$, Fig.~\ref{ip_focal}E), caustic focusing produces a near-line focus with $\omega_{0x} = 1.7~\mu$m and $\omega_{0y} = 11.2~\mu$m (Fig.~\ref{ip_focal}F), giving a substantially higher local intensity at the rear surface.

With increasing pulse energy, plasma-induced defocusing broadens the rear focal spot for both geometries~\cite{Singh2026Caustic}. The focal width $2\omega_{0x}$ at $\mathrm{I.P.} = 0.71$ nevertheless remains smaller than at $\mathrm{I.P.} = 0$ throughout the measured energy range, so caustic focusing retains its advantage even where defocusing is significant. Further details of the focal profiles and their energy dependence are given in Ref.~\cite{Singh2026Caustic}.

\subsection{Coupled control of the dominant interface by pulse energy and caustic focusing}
\label{sec:energy_time_series}

\begin{figure}[h!]
\centering
\includegraphics[width=0.8\linewidth]{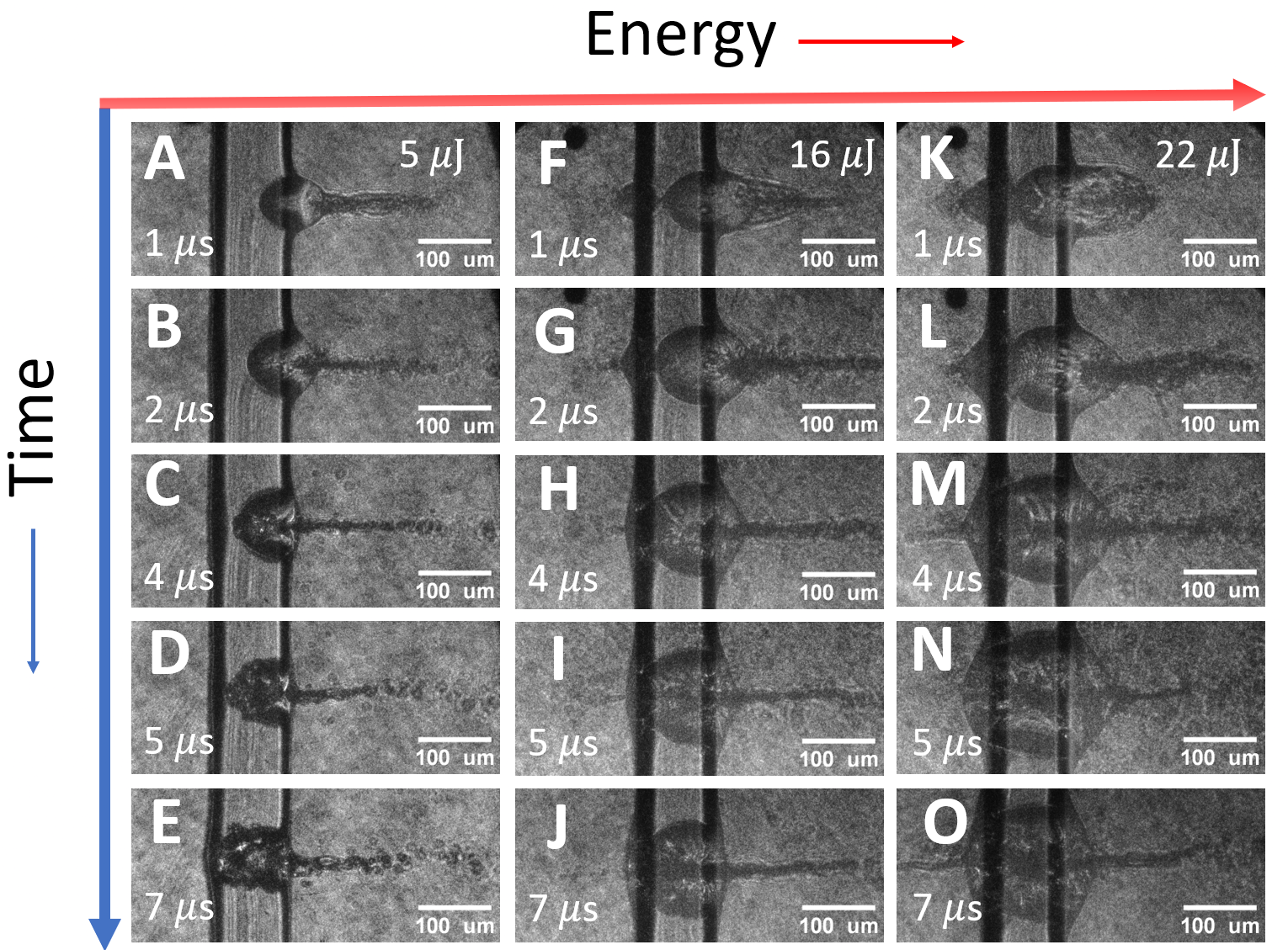}
\caption{\textbf{Time--energy map of the interfacial response under off-centre irradiation ($\mathrm{I.P.} \approx 0.85 \pm 0.1$).} Shadowgraphs recorded 1--7~$\mu$s after the pulse for pulse energies of 5~$\mu$J (\textbf{A--E}), 16~$\mu$J (\textbf{F--J}), and 22~$\mu$J (\textbf{K--O}). Cavitation and jetting are dominated by the rear interface at all three energies.}
\label{energy_curvature_ablation}
\end{figure}

\begin{figure}[h!]
\centering
\includegraphics[width=\linewidth]{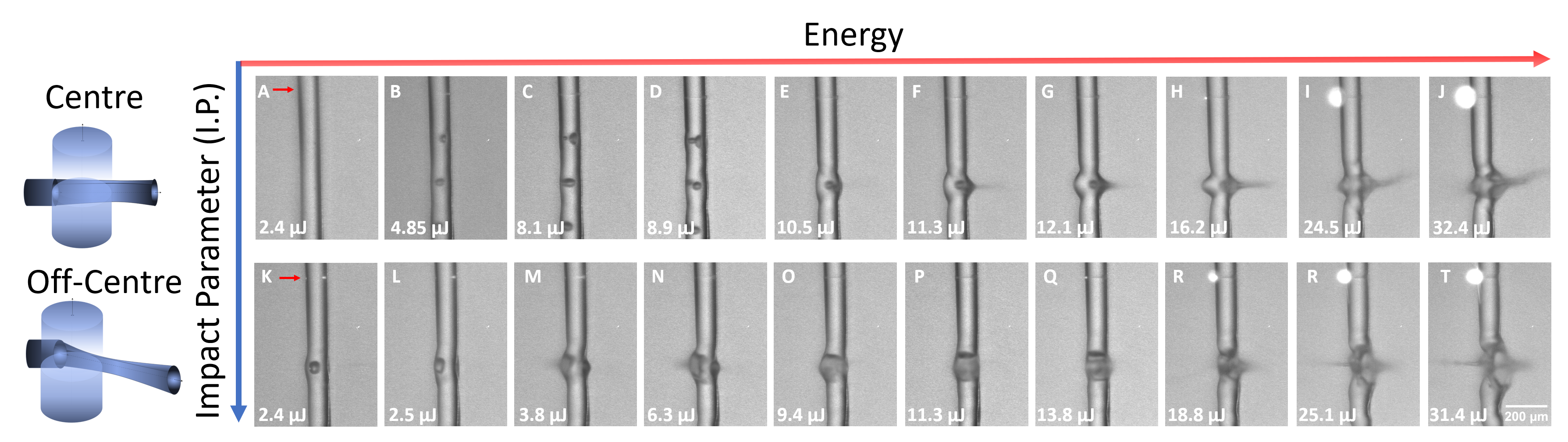}
\caption{\textbf{Energy dependence of the interfacial response in the two irradiation geometries}, recorded $\approx 40~\mu$s after the pulse. \textbf{(A--J)} Centred irradiation ($\mathrm{I.P.} = 0$), 2.4--32.4~$\mu$J. \textbf{(K--T)} Off-centre irradiation ($\mathrm{I.P.} = 0.71$), 2.4--31.4~$\mu$J. Jetting appears from 11.3~$\mu$J (F) under centred irradiation and from 3.8~$\mu$J (M) under off-centre irradiation, where the response is balanced at 13.8~$\mu$J (Q) and front-dominated from 18.8~$\mu$J (R--T).}
\label{energy_ip_curvature}
\end{figure}

The interface at which cavitation dominates, and therefore the direction of the recoil and of the resulting curvature, is not set by a single parameter. It emerges from the interplay of two coupled controls. The impact parameter sets, through the cylindrical lensing of the jet, which interface the light favours: centred irradiation concentrates the irradiance at the front interface, whereas off-centre irradiation lowers it there and focuses the transmitted light toward the rear. The pulse energy then shifts this balance, because breakdown ~\cite{Vogel2005,Optical_breakdown_Vogel} and plasma absorption along the beam path increasingly favour the front interface as the energy rises (main-text Fig.~\ref{fig:frontbackintensity}A). Which interface ruptures first, which one jets most strongly, and hence in which direction the column recoils and bends, is decided by the combination of the two. We map this coupling with two complementary datasets: a time--energy series at fixed impact parameter under strong caustic focusing (Supplementary Fig.~\ref{energy_curvature_ablation}), and an energy series at fixed delay for two impact parameters (Supplementary Fig.~\ref{energy_ip_curvature}).

Under caustic focusing ($\mathrm{I.P.} \approx 0.85$, Supplementary Fig.~\ref{energy_curvature_ablation}), the rear interface dominates at every energy studied. At 5~$\mu$J, strong jetting emerges from the rear interface by 1~$\mu$s, while the front interface remains unperturbed (A). By 2~$\mu$s the jetting slows, and the ruptured rear region recoils (B), leaving a deformed rear interface by 7~$\mu$s as the column is set in motion (C--E). At 16~$\mu$J, weak ablation and cavitation also appear at the front interface, but the rear jetting remains much stronger (F--J). At 22~$\mu$J, cavitation is seeded at both interfaces, and the rear still dominates (K--O). In all three cases the jetting subsides within a few microseconds and the net recoil, directed toward the front, drives the subsequent slow motion of the column.

The energy series at fixed delay (Supplementary Fig.~\ref{energy_ip_curvature}) compares the two geometries directly. Under centred irradiation, bubbles form at low energies without rupturing the interfaces or breaking the axisymmetry of the jet (A--E). Jetting first appears at 11.3~$\mu$J (F) and strengthens with energy, and by 32.4~$\mu$J pronounced front-side jetting accompanies a clear rearward displacement of the column (G--J). Under off-centre irradiation, caustic focusing activates the rear interface at much lower energies: rupture and jetting appear from 3.8~$\mu$J (M), compared with 11.3~$\mu$J in the centred case. As the energy increases, the front interface catches up. At 13.8~$\mu$J the responses of the two interfaces balance and the segment remains straight (Q), and from 18.8~$\mu$J the front dominates and the asymmetry reverses (R--T).

Taken together, the two datasets show the coupling directly. The energy at which dominance passes from the rear to the front interface depends on the impact parameter. Under centred irradiation the front dominates wherever jetting occurs. At $\mathrm{I.P.} = 0.71$ the rear dominates at low energy and the balance point lies near 13.8~$\mu$J, close to the measured front--rear intensity crossover (main-text Fig.~\ref{fig:frontbackintensity}A). At $\mathrm{I.P.} \approx 0.85$, where oblique incidence lowers the front irradiance further, rear dominance persists to at least 22~$\mu$J. Increasing the impact parameter therefore shifts the balance point to higher energies. The impact parameter and the pulse energy thus act as a single coupled control space, in which their combination selects the dominant interface and, with it, the direction of the recoil and of the curvature.

\subsection{Balanced front--rear deposition: a straight-segment control}
\label{sec:balance_control}

\begin{figure}[h!]
    \centering
    \includegraphics[width=0.8\linewidth]{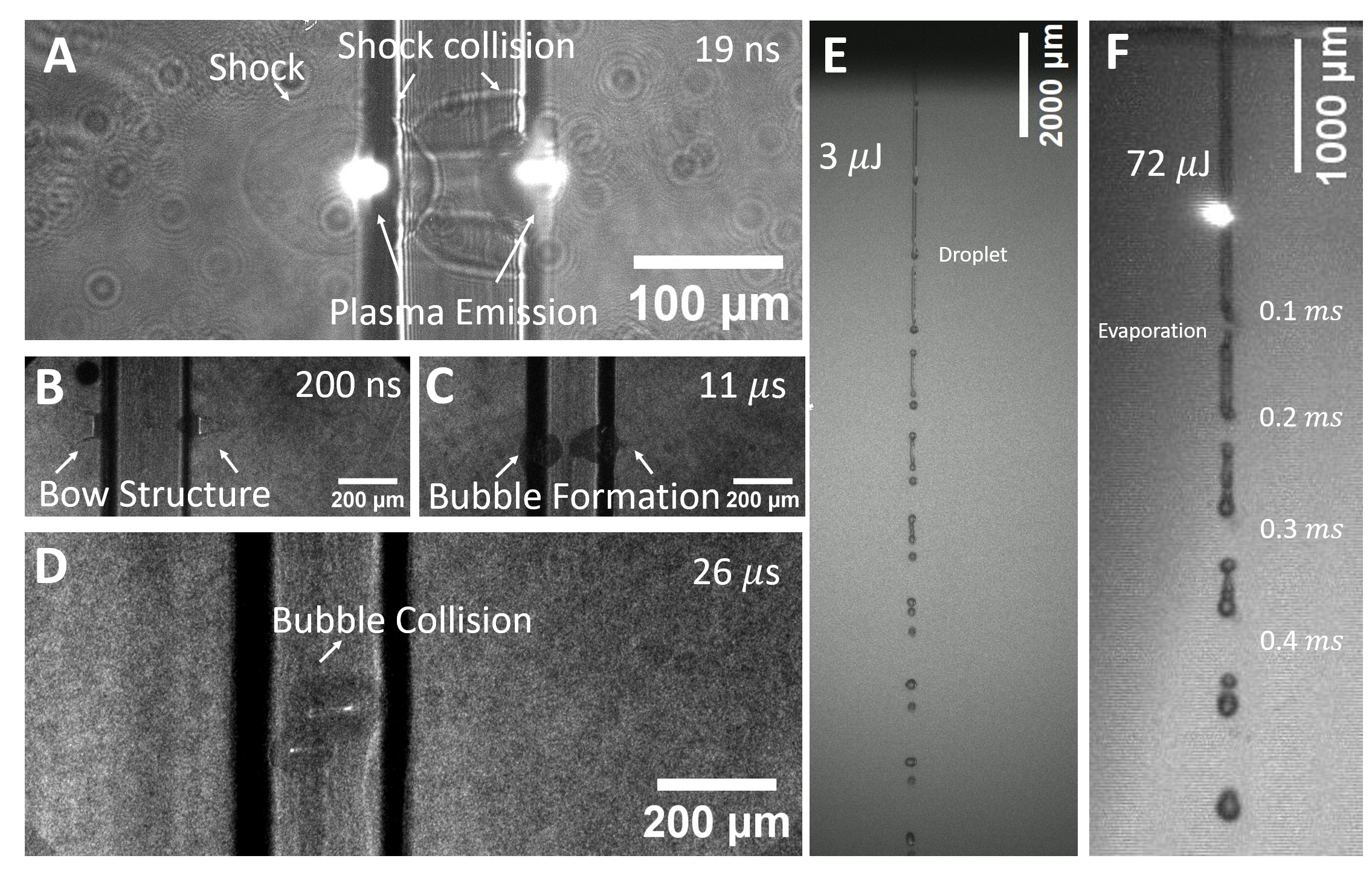}
    \caption{\textbf{Balanced front--rear deposition produces straight segments.}
    \textbf{(A)} 19\,ns: plasma at both interfaces, connected by a plasma channel across the jet, with shocks launched from both sides colliding inside the jet.
    \textbf{(B)} 200\,ns, $10.4~\mu$J: bow-shaped structures of comparable size at both interfaces.
    \textbf{(C)} 11\,$\mu$s: cavitation bubbles at both interfaces.
    \textbf{(D)} 26\,$\mu$s: the bubbles from the two interfaces collide inside the jet.
    \textbf{(E)} $3~\mu$J: weak breakdown seeds pinch-off into straight segments that contract into drops.
    \textbf{(F)} $72~\mu$J: strong but balanced interaction leaves straight segments that retract and collapse into drops by $\approx 0.4$~ms.}
    \label{shock_bubble}
\end{figure}

The curvature reported in the main text arises from an imbalance between the cavitation recoils at the front and rear interfaces. When the two interfaces are driven equally, no curvature forms. Under balanced conditions, plasma forms at both interfaces with comparable brightness, connected by a plasma channel across the jet, and shocks launched from both sides collide inside the jet (Fig.~\ref{shock_bubble}A). The ablation that follows is equally balanced, with bow-shaped structures ~\cite{Vogel2003} of comparable size at both interfaces by 200~ns (Fig.~\ref{shock_bubble}B). Cavitation bubbles then nucleate at both interfaces (Fig.~\ref{shock_bubble}C), and their recoils drive liquid inward from both sides until the bubbles collide inside the jet at $\approx 26~\mu$s (Fig.~\ref{shock_bubble}D). The opposing recoils cancel, leaving no net transverse momentum and no preferred bending direction.

Two limits confirm that bending requires an asymmetry rather than cavitation alone. At $3~\mu$J (Fig.~\ref{shock_bubble}E), breakdown is too weak to break the axisymmetry of the jet. It only seeds pinch-off, and the resulting straight segments retract from both ends and contract into drops. At $72~\mu$J (Fig.~\ref{shock_bubble}F), the interaction is strong, with intense plasma formation, cavitation, and jetting, but it remains balanced between the two interfaces. The segments again remain straight and collapse into drops through end retraction by $\approx 0.4$~ms, the dynamics used to measure the retraction speed in Supplementary Section~\ref{sec:bent_segment}. In both limits, once the laser has acted, a straight segment evolves only by end retraction, and curvature appears only when one interface dominates the deposition.

\subsection{Curvature extraction and geometry of bent segments}
\label{sec:curvature_method}
\label{sec:bent_segment}

\begin{figure}[h!]
    \centering
    \includegraphics[width=\linewidth]{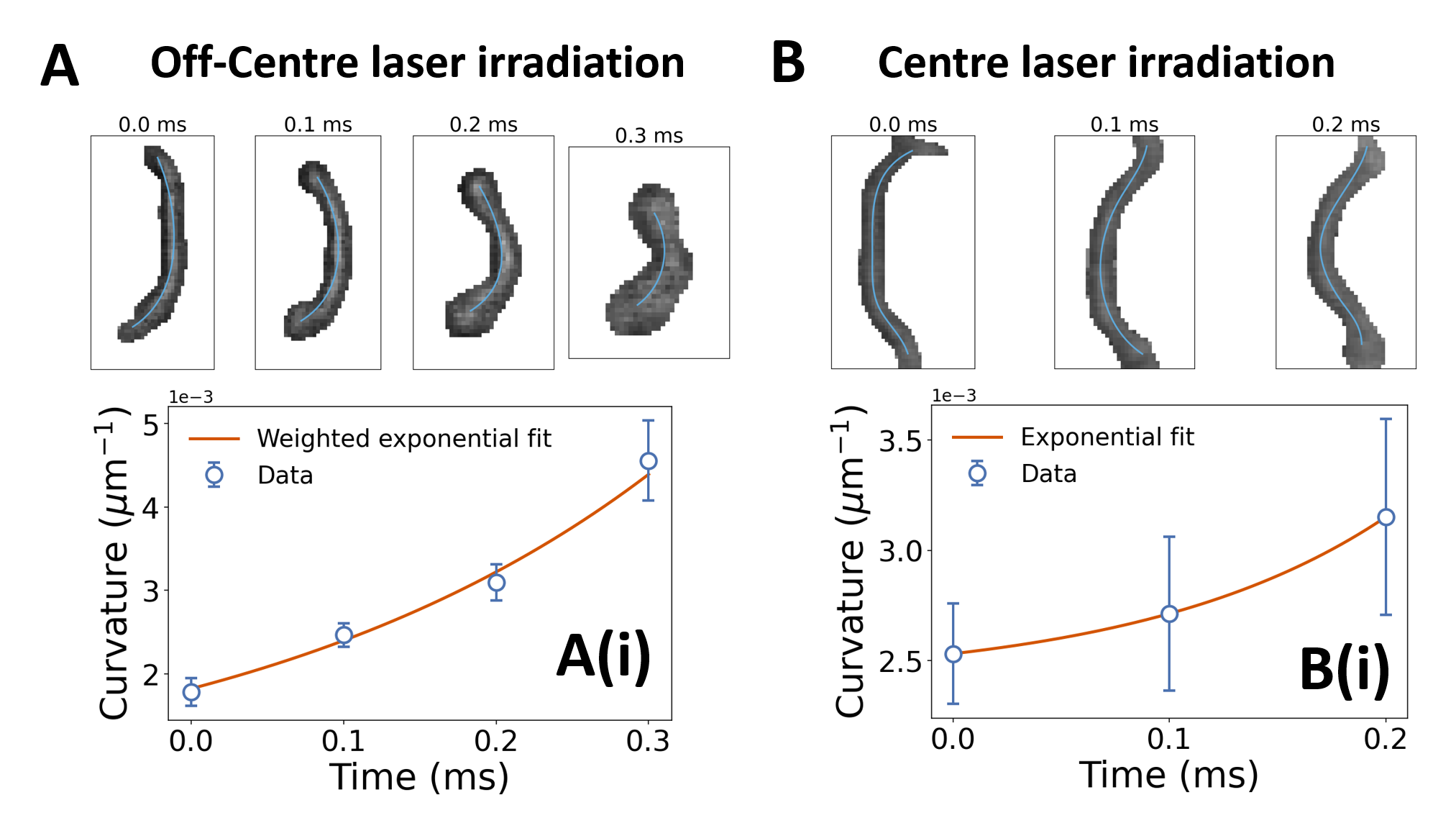}
    \caption{\textbf{Morphological evolution of laser-bent segments.}
    \textbf{(A)} Off-centre irradiation produces one curvature sign, whose magnitude increases with time.
    \textbf{(B)} Centred irradiation produces the opposite curvature sign.}
    \label{fig:jetcurvature}
\end{figure}

The reported curvatures characterize finite segments bounded by successive cavitation sites. Each segment centreline was fitted with a smoothing spline, $(x(q),y(q))$, and its local curvature calculated as
\begin{equation}
\kappa_{\mathrm{loc}} =
\frac{|x'(q)y''(q)-y'(q)x''(q)|}
{[x'(q)^2+y'(q)^2]^{3/2}},
\label{eq:curvature}
\end{equation}
where primes denote derivatives with respect to $q$. The segment curvature is the centreline average of $\kappa_{\mathrm{loc}}$, with its sign assigned according to the bend direction: positive along the beam and negative toward the laser. Figure~\ref{fig:jetcurvature} shows representative morphologies for the two irradiation geometries in Fig.~\ref{fig:jetcurvaturea}.

\paragraph{Surface energy of a bent segment.}
For a cylindrical segment of radius $R$, whose centreline forms an arc of radius $R_c$ and angle $\Delta\theta$, the surface area and volume are
\begin{equation}
A=2\pi RL,\qquad V=\pi R^2L,
\label{eq:bent_geometry}
\end{equation}
where $L=R_c\Delta\theta$ is the centreline length. These quantities are independent of $R_c$ and identical to those of a straight segment of equal length and radius. Thus, bending at fixed length does not change the surface energy, $\sigma A$, and surface tension provides no energetic preference for a straight configuration.

\paragraph{Laplace pressure of a bent segment.}
Although the total surface energy is unchanged, bending introduces a variation in the local interfacial curvature. The principal curvatures are $\kappa_1=1/R$ and
\begin{equation}
\kappa_2=\frac{\cos\phi}{R_c+R\cos\phi},
\label{eq:principal_curvatures}
\end{equation}
where $\phi$ is measured from the outer side of the bend. The resulting Young--Laplace pressure difference between the outer and inner sides is approximately $2\sigma/R_c$ for $R\ll R_c$. For $R=50~\mu$m and $R_c\approx0.9$~mm, this difference is $\approx160$~Pa, compared with a mean capillary pressure $\sigma/R\approx1.4$~kPa. The circumferential average of $\kappa_2$ is zero.

\paragraph{End retraction.}
Surface tension also drives retraction of the free ends. The characteristic capillary--inertial velocity scale is
\begin{equation}
c=\sqrt{\frac{\sigma}{\rho R}}\approx1.2~\mathrm{m\,s^{-1}}.
\end{equation}
We measured this process for a straight segment of initial length $L_0\approx540~\mu$m ($L_0/D\approx5.4$), which collapses into a single droplet within $\approx0.4$~ms. Volume conservation gives the final droplet diameter
\begin{equation}
d=\left(\frac{3D^2L_0}{2}\right)^{1/3}
\approx200~\mu\mathrm{m}.
\end{equation}
The corresponding retraction distance is $L_0-d\approx340~\mu$m, giving an average retraction speed of $\approx0.4~\mathrm{m\,s^{-1}}$ per end, approximately one-third of the capillary--inertial velocity scale.

\paragraph{Evolution of bent segments.}
Two capillary processes may contribute to the evolution: end retraction and the pressure gradient across the bent interface. End retraction acts on both straight and bent segments, whereas the outer-to-inner Laplace pressure difference could additionally drive bending-induced deformation. Their contributions cannot be isolated experimentally because curvature and liquid removal are coupled: stronger laser coupling produces greater ablation and jetting, changing the residual segment length and volume. These quantities therefore differ between straight and bent segments and between the two bending directions. Nevertheless, the comparable retraction timescales observed for straight segments and the evolution of bent segments are consistent with end retraction dominating the observed collapse. We find no clear evidence of additional pressure-driven folding.

\subsection{Supplementary Videos}

\noindent\textbf{Supplementary Video 1.} Laser-induced seeding at 19~ns after irradiation, illustrating the effect of impact parameter on initial plasma formation and shock-wave generation.

\noindent\textbf{Supplementary Video 2.} Laser-induced plume evolution at 0.9~$\mu$s after irradiation, illustrating the effect of impact parameter on plume direction.

\noindent\textbf{Supplementary Video 3.} Zoomed-out view of laser-induced liquid curvature dynamics at 40~$\mu$J, from initial perturbation (14~ns) to 0.1~ms after irradiation.

\noindent\textbf{Supplementary Video 4.} Effect of impact parameter on the direction of laser-induced curvature at 60~$\mu$J.

\noindent\textbf{Supplementary Video 5.} Effect of input laser intensity on the direction of laser-induced curvature for a $50~\mu\mathrm{m}$-diameter jet.

\noindent\textbf{Supplementary Video 6.}  Effect of input laser intensity on the direction of laser-induced curvature for a $100~\mu\mathrm{m}$-diameter jet.



\bibliographystyle{naturemag}
\bibliography{references}

\end{document}